\documentclass[twocolumn,prb,amsmath,amssymb,showpacs,superscriptaddress,floatfix]{revtex4}

\usepackage{graphicx}

\DeclareMathOperator{\Tr}{Tr}
\begin{document}

\title{Bell inequalities and entanglement in solid state devices}

\author{Nikolai~M. Chtchelkatchev}
\email{nms@landau.ac.ru}\affiliation{L.D.\ Landau Institute for Theoretical Physics RAS, 117940 Moscow, Russia}
\affiliation{Centre de Physique Th\'eorique, Universit\'e de la M\'editerran\'ee, Case 907, 13288 Marseille, France}

\author{Gianni Blatter}\affiliation{Theoretische Physik, ETH-H\"onggerberg, CH-8093 Z\"urich, Switzerland}

\author{Gordey~B.~Lesovik}\affiliation{L.D.\ Landau Institute
for Theoretical Physics RAS, 117940 Moscow, Russia}\affiliation{Theoretische Physik, ETH-H\"onggerberg, CH-8093
Z\"urich, Switzerland}

\author{Thierry Martin}
\affiliation{Centre de Physique Th\'eorique, Universit\'e de la M\'editerran\'ee, Case 907, 13288 Marseille, France}

%\date{24 February 2002}

\begin{abstract}
Bell-inequality checks constitute a probe of entanglement --- given a source of entangled particles, their violation
are a signature of the non-local nature of quantum mechanics. Here, we study a solid state device producing pairs of
entangled electrons, a superconductor emitting Cooper pairs properly split into the two arms of a normal-metallic fork
with the help of appropriate filters. We formulate Bell-type inequalities in terms of current-current
cross-correlators, the natural quantities measured in mesoscopic physics; their violation provides evidence that this
device indeed is a source of entangled electrons.
\end{abstract}


\maketitle

Entanglement is a defining feature  of quantum mechanical systems \cite{Schrodinger,EPR,Menskii,book:q_information}
with important new applications in the emerging fields of quantum information theory,\cite{Zeilenger} quantum
computation,\cite{Steane} quantum cryptography,\cite{Ekert} and quantum teleportation.\cite{Bennet} Many examples of
entangled systems can be found in nature, but only in few cases can entanglement be probed and used in applications. So
far, much attention has been focused on the preparation and investigation of entangled photons
\cite{Aspect,book:Mandel_Wolf} and, more recently, of entangled atoms,\cite{Cirac,Rowe} while other studies use
elementary particles (kaons) \cite{Bertlmann} and electrons.\cite{DiVincenzo} Bell inequality (BI) \cite{Bell} checks
have become the accepted method to test entanglement:\cite{Werner,Mermin} their violation in experiments with particle
pairs indicates that there are nonlocal correlations between these particles as predicted by quantum mechanics which no
local hidden variable theory can explain.\cite{Bell}

Quasi-particles in solid state devices are promising candidates as carriers of quantum information. Recent
investigations provide strong evidence that electron spins in a semiconductor show unusually long dephasing times
approaching microseconds; furthermore, they can be transported phase coherently over distances exceeding 100
$\mu$m.\cite{Kikkawa} Several proposals how to create an Einstein-Podolsky-Rosen \cite{EPR} (EPR) pair of electrons in
solid-state systems have been made recently; one of these is to use a superconductor as a source of entangled beams of
electrons.\cite{Lesovik1,Loss} At first glance, the possibility of performing BI checks in solid state systems may seem
to be a naive generalization \cite{Kawabata,Ionicioiu} of the corresponding tests with
photons.\cite{Aspect,book:Mandel_Wolf} But in the case of photons, the BIs have been tested using photodetectors
measuring coincidence rates (the probability that two photons enter the detectors nearly simultaneously
\cite{Aspect,book:Mandel_Wolf}). Counting quasi-particles one-by-one (as photodetectors do in quantum optics
\cite{book:Mandel_Wolf}) is difficult to achieve in solid-state systems where currents and current-current correlators,
in particular noise, are the natural observables in a stationary regime.\cite{Blanter-Buttiker} Here, the BIs are
re-formulated in terms of current-current cross-correlators (noise) and a practical implementation of BIs as a test of
quasi-particle entanglement produced via a hybrid superconductor--normal-metal source \cite{Lesovik1,Loss} is
discussed.\cite{Werner,Popesku}
\begin{figure}
\includegraphics[width=80mm]{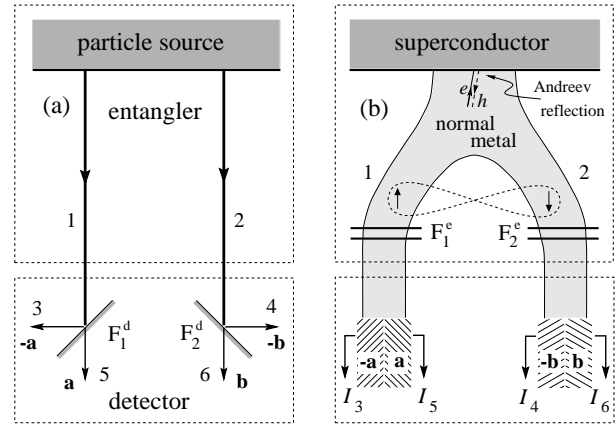}
\caption{\label{fig_1} Schematic setup (a) and solid state implementation (b) for the measurement of Bell inequalities:
a source emits particles into leads 1 and 2. The detector measures the correlation between beams labelled with odd and
even numbers. The filters F$^{\rm d}_{1(2)}$ select the spin: particles with polarization along the direction $\bf a$
are transmitted through filter F$^{\rm d}_1$ into lead 5, while the other electrons are channelled into lead 3
(and similar for F$^{\rm d}_2$).
The solid state implementation (b) involves a superconducting source emitting Cooper pairs into the
leads. The filters F$^{\rm e}_{1,2}$ (realized, e.g., via Fabry-Perot double barrier structures or quantum
dots) prevent the Cooper pairs from decaying into a single lead. Ferromagnets play the
role of the filters F$^{\rm d}_{1(2)}$ in the detector (here used in a Stern-Gerlach type geometry); they are
transparent for electrons with spin aligned along their magnetization.}
\end{figure}

Consider a source [Fig.~1(a)] injecting quasi-particles into two arms labelled by indices 1 and 2. The detector
includes two filters F$^{\rm d}_{1(2)}$ selecting electrons by spin; the filter F$^{\rm d}_1$ transmits electrons
spin-polarized along the direction $\bf a$ into lead 5 and deflects electrons with the opposite polarization into lead
3 (and similar for filter F$^{\rm d}_2$ with direction $\bf b$). The detector thus measures cross-correlations of
(spin-)currents between the leads; a violation of BIs provides evidence for nonlocal spin-correlations between the
quasi-particle beams 1 and 2.

We formulate the BIs in terms of current-current correlators: assuming separability and locality \cite{Bell,Werner} (no
entanglement, only local correlations are allowed) the density matrix of the source/detector system describing joint
events in the leads $\alpha,\beta$ is given by
\begin{gather}
   \label{rho}
   \rho=\int d\lambda f(\lambda)
   \rho_\alpha(\lambda)\otimes\rho_\beta(\lambda),
\end{gather}
where the lead index $\alpha$ is even and $\beta$ is odd (or vice-versa); the distribution function $f(\lambda)$
(positive and normalized to unity) describes the `hidden variable' $\lambda$. The Hermitian operators
$\rho_\alpha(\lambda)$ satisfy the standard axioms of density matrices. For identical particles the assumption
\eqref{rho} implies that Bose and Fermi correlations between leads with odd and even indices are neglected.

Consider the Heisenberg operator of the current $I_\alpha(t)$ in lead $\alpha=1,\ldots,6$ (see Fig.~1) and the
associated particle number operator $N_\alpha(t,\tau) =\int_t^{t+\tau} dt^\prime\, I_\alpha (t^\prime)$ describing the
charge going through a cross-section of lead $\alpha$ during the time interval $[t,t+\tau]$. We define the
particle-number correlators $\langle N_\alpha(t,\tau) N_\beta(t,\tau)\rangle_\rho=\int d\lambda f(\lambda) \langle
N_\alpha(t,\tau)\rangle_\lambda\langle N_\beta(t,\tau)\rangle_\lambda$ (with indices $\alpha/\beta$ odd/even or
even/odd), where $\langle N_\alpha(t,\tau) \rangle_\lambda \equiv \Tr[\rho_\alpha(\lambda)N_\alpha(t,\tau)]$ and
$\langle\ldots\rangle_\rho \equiv \Tr[\rho\ldots]$. The average $\langle N_\alpha(t,\tau) \rangle_\lambda$ depends on
the state of the system in the interval $[t,t+\tau]$; in general $\langle N_\alpha(t_1,\tau) \rangle_\lambda
\neq \langle N_\alpha(t_2,\tau)\rangle_\lambda$, where $t_1\neq t_2$. For later convenience we introduce
the average over large time periods in addition to averaging over $\lambda$, e.g.,
\begin{gather}
   \label{time averaging} \langle
   N_\alpha(\tau)N_\beta(\tau)\rangle
   \equiv \frac 1{2T}\int_{-T}^T dt\langle
   N_\alpha(t,\tau)N_\beta(t,\tau)\rangle_\rho,
\end{gather}
where $T/\tau\to \infty$ (a similar definition applies to $\langle N_\alpha(\tau)\rangle$). Finally, we define the
particle number fluctuations $\delta N_\alpha(t,\tau)\equiv N_\alpha(t,\tau)-\langle N_\alpha(\tau)\rangle$.

The derivation of the Bell inequality is based on the following lemma: let $x,x^\prime,y,y^\prime,X,Y$ be real numbers
such that $|x/X|$, $|x^\prime/X|$,  $|y/Y|$, and $|y^\prime/Y|$ do not exceed unity, then the following inequality
holds:\cite{lemma}
\begin{gather}
   \label{lemma}-2XY\leq xy-xy^\prime+x^\prime
   y+x^\prime y^\prime\leq 2XY.
\end{gather}
Lemma \eqref{lemma} is applied to our system with
\begin{subequations}
   \begin{eqnarray}
   x&=& \langle N_5(t,\tau)\rangle_\lambda
   -\langle N_{3}(t,\tau)\rangle_\lambda, \\
   x^\prime& =&
   \langle N_{5^\prime}(t,\tau)\rangle_\lambda -\langle
   N_{3^\prime}(t,\tau)\rangle_\lambda,\label{xy}\\
   y&=&\langle
   N_6(t,\tau)\rangle_\lambda-\langle
   N_4(t,\tau)\rangle_\lambda, \\
   y^\prime&=&\langle
   N_{6^\prime}(t,\tau)\rangle_\lambda-\langle
   N_{4^\prime}(t,\tau)\rangle_\lambda,
   \end{eqnarray}
\end{subequations}
where the `prime' indicates a different direction of spin-selection in the detector's filter (e.g., let $\bf a$ denote
the direction of the electron spins in lead 5 ($-\bf a$ in lead 3), then the subscript $5^\prime$ in Eq.~\eqref{xy}
refers to electron spins in lead 5 polarized along $\bf a^\prime$ (along $-\bf a^\prime$ in the lead 3). The quantities
$X,Y$ are defined as
\begin{subequations}
   \begin{eqnarray}
   X= \langle  N_5(t,\tau)\rangle_\lambda
   +\langle N_{3}(t,\tau)\rangle_\lambda=\nonumber\\
   \langle N_{5^\prime}(t,\tau)\rangle_\lambda
   +\langle N_{3^\prime}(t,\tau)\rangle_\lambda= \langle
   N_1(t,\tau)\rangle_\lambda,\label{X} \\
   Y= \langle N_6(t,\tau)\rangle_\lambda
   +\langle N_{4}(t,\tau)\rangle_\lambda=\nonumber\\
   \langle N_{6^\prime}(t,\tau)\rangle_\lambda
   +\langle N_{4^\prime}(t,\tau)\rangle_\lambda=\langle
   N_2(t,\tau)\rangle_\lambda; \label{Y}
   \end{eqnarray}
\end{subequations}
the equalities \eqref{X} and \eqref{Y} follow from particle number conservation. All terms in \eqref{X} and \eqref{Y}
have the same sign, hence $|x/X|\leq 1$ and $|y/Y|\leq 1$.

The Bell-inequality follows from (\ref{lemma}) after averaging
over both time $t$ [see Eq.~\ref{time averaging}] and $\lambda$,
\begin{gather}
   \label{Bell}
   |G({\bf a,b})-G({\bf a,b^\prime})+ G({\bf a^\prime,b})
   +G({\bf a^\prime,b^\prime})|\leq 2,
\end{gather}
where
\begin{gather}
   G({\bf a,b})
   =\frac{\langle(N_5(\tau)-N_3(\tau))(N_6(\tau)
   -N_4(\tau))\rangle}{\langle(N_5(\tau)+
   N_3(\tau))(N_6(\tau)+N_4(\tau))\rangle} \nonumber,
\end{gather}
and with $\bf a, b$ the polarizations of the filters F$^{\rm
d}_{1(2)}$.

At this point, the number averages and correlators in \eqref{Bell} need to be related to measurable quantities, current
averages and current noise; this step requires to perform the time averaging introduced in \eqref{time averaging} and
implemented in \eqref{Bell}. The correlator $\langle N_\alpha(\tau) N_\beta(\tau)\rangle$ includes both reducible and
irreducible parts. As demonstrated below, the Bell inequality \eqref{Bell} can be violated if the irreducible part of
the correlator is of the order of (or larger) than the reducible part. The irreducible correlator $\langle\delta
N_\alpha(\tau)\delta N_\beta(\tau) \rangle$ can be expressed through the noise power $S_{\alpha\beta}(\omega)=\int
d\tau e^{i\omega \tau} \langle\delta I_\alpha(\tau)\delta I_\beta(0)\rangle$,
\begin{gather}
   \label{ir_cor}
   \langle\delta N_\alpha(\tau)\delta N_\beta(\tau)\rangle
   =\int_{-\infty}^\infty \frac{d\omega}{2\pi}
   S_{\alpha\beta}(\omega)
   \frac{4\sin^2\left(\omega \tau/2\right)}{\omega^2}.
\end{gather}
In the limit of large times, $\sin^2(\omega \tau/2) /(\omega/2)^2
\to 2\pi \tau\delta(\omega)$, and therefore
\begin{gather}
   \label{cor}
   \langle N_\alpha(\tau)N_\beta(\tau)\rangle\approx \langle
   I_\alpha\rangle \langle I_\beta\rangle \tau^2+\tau S_{\alpha\beta},
\end{gather}
where $\langle I_\alpha\rangle$ is the average current in the lead $\alpha$ and $S_{\alpha\beta}$ denotes the shot
noise. In reality, the noise power diverges as $1/\omega$ when $\omega\to 0$, but this singular behavior starts from
very small $\omega$ ($\omega\ll \omega_{\rm fl}\sim 10^{-3}$s$^{-1}$).\cite{Imry} At frequencies $\omega_{\rm
fl}\ll\omega\ll\omega_{0}$ the noise power is nearly constant (see, e.g., Ref. \onlinecite{Blanter-Buttiker}). The
upper boundary $\omega_0$ of the frequency domain depends on the voltage $V$ of the terminals $3-6$ (the particle
source is grounded), on the characteristic time of electron flight $\tau_{\rm tr}$ between these terminals, and the
widths $\Gamma_{1(2)}$ of the filters F$^{\rm e}_{1,2}$ which each have a resonant energy $\pm E_0$ [see Fig.~1(b)],
$\omega_0=\min(|V|;\Gamma_{1(2)}; \tau_{\rm tr}^{-1})$. Thus \eqref{ir_cor} implies \eqref{cor} if
$\omega_0^{-1}\ll\tau\ll\omega_{\rm fl}^{-1} $ [we assume a temperature $T<\omega_0$]. Using \eqref{Bell} and
\eqref{cor} we find
\begin{subequations}
   \begin{eqnarray}
   \label{Bell_noise}
   |F({\bf a,b})-F({\bf a,b^\prime})+ F({\bf a^\prime,b})
   +F({\bf a^\prime,b^\prime})|\leq 2,\\
   F({\bf a,b})=\frac{S_{56}-S_{54}-S_{36}+S_{34}
   +\Lambda_-}{S_{56}+S_{54}+S_{36}+S_{34}+\Lambda_+},
   \label{Bell_noise_F}
   \end{eqnarray}
\end{subequations}
where $\Lambda_\pm=\tau(\langle I_5\rangle \pm\langle I_3\rangle)(\langle I_6\rangle \pm\langle I_4\rangle)$. The Bell
inequality (\ref{Bell_noise}) is the expression to be tested in the experiment; as implied by \eqref{Bell_noise_F} its
violation requires the dominance of the irreducible particle-particle correlator encoded in the shot noise
$|S_{\alpha\beta}|\gtrsim |\Lambda_\pm|$.

Below we discuss the violation of the above Bell inequality in mesoscopic systems. As a general rule, the violation of
\eqref{Bell_noise} implies that the assumption \eqref{rho} does not hold and the correlations are non-classical. In
this situation, particles injected by the source S into leads 1 and 2 (see Fig.~1) are entangled (if the system is in a
pure state, the entanglement implies that its wave function cannot be reduced to a product of wave functions
corresponding to particles in leads 1 and 2).

Consider now the solid-state analog of the Bell-device as sketched in Fig.~1(b) where the particle source is  a
superconductor (S). Two normal-metal leads 1 and 2 are attached in a fork geometry to the particle source
\cite{Lesovik1,Loss} and the energy- or charge-selective filters F$^{\rm e}_{1,2}$ enforce the splitting of the
injected pairs. Ferromagnetic filters play the role of spin-selective beam-splitters F$^{\rm d}_{1,2}$ in the detector
(suitable filters F$^{\rm d}_{1(2)}$ can be constructed with the help of ferromagnets,\cite{Lesovik1} quantum
dots,\cite{Recher} and hybrid superconductor--normal-metal--ferromagnet structures \cite{Huertas-Hernando}): e.g.,
quasi-particles injected into lead 1 ($I_1$) and spin-polarized along the magnetization $\bf a$ enter the ferromagnet 5
and contribute to the current $I_5$, while quasi-particles with the opposite polarization contribute to the current
$I_{3}$, see Fig.~1(b). The appropriate choice of voltages between the leads and the source fixes the directions of the
currents in agreement with Fig.~1(a). The test of the Bell inequality \eqref{Bell_noise} requires information about the
dependence of the noise on the mutual orientations of the magnetizations $\pm{\bf a}$ and $\pm{\bf b}$ of the
ferromagnetic spin-filters (see Fig.~1(b)).

The noise power is calculated using scattering
theory.\cite{Lesovik2,Blanter-Buttiker} Normal leads are labelled with Greek letters $\alpha,\beta,\ldots$, electron
(hole) charges are denoted by $q_a$, where $a=e(h)$, e.g., $q_e=-1$, $q_h=1$. If $C_\alpha$ is the number of channels
in the lead $\alpha$, then the amplitude for scattering of a quasi-particle $a$ from the lead $\alpha$ into a
quasi-particle $b$ in the lead $\beta$ is given by the scattering matrix $s_{ab}^{\alpha\beta}$ (of dimension $C_\beta
\times C_\alpha$). The expression for the noise power takes the form \cite{Anatram-Datta}
\begin{multline}
   \label{noise}
   S_{\alpha\beta}=\frac{e^2}{h}\int_{0}^\infty dE
   \sum_{\gamma,\delta;a,b,c,d}f_{\gamma,a}(1-f_{\delta,b})\\
   \Tr[(s^\dag)^{\gamma\alpha}_{ac}q_c s^{\alpha\delta}_{cb}
   (s^\dag)^{\delta\beta}_{bd}q_d s_{da}^{\beta\gamma} ],
\end{multline}
where the energy is measured with respect to the electrochemical potential of the particle source; $V_\alpha$ is the
voltage in lead $\alpha$, $f_{\alpha,a}=1/(\exp\{(E-q_aV_\alpha)/T\}+1)$, and the trace is taken over all channel
degrees of freedom. We assume weak coupling between the superconductor and the leads 1 and 2 with electrons entering
the superconductor through a tunnel barriers with normal (dimensionless) conductances $g_{1(2)} \ll 1$, hence
$\Lambda_\pm\sim \tau (\omega_1g_{1}g_2)^2$,\cite{note} where $\omega_1=\min(|V|;\Gamma_{1(2)})$, $|V|<\Delta$. It
follows from \eqref{noise} that $S_{\alpha\beta}\sim \omega_{1}g_{1}g_2 $. Thus the condition $\omega_{1}\tau g_{1}g_2
\ll 1$ (i.e., no more than one quasi-particle pair can be detected during the measurement time $\tau$) allows to drop
$\Lambda_\pm$ in \eqref{Bell_noise_F}. Eq.~(\ref{Bell_noise}) becomes  the nonlocality criterium if there is no
electron exchange between the leads 1 and 2 during  the measurement time $\tau$, requiring $\tau_{\rm tr}^{-1}\tau
g_{1}g_2 \ll 1$.\cite{explanation} The two conditions can be written as $\omega_{0}\tau g_{1}g_2 \ll 1$; the
corresponding BI violation is discussed below.

The matrix under the trace in \eqref{noise} depends on ${\bf a} \cdot \vec\sigma$, ${\bf b}\cdot\vec\sigma$; making use
of the relation
\begin{multline}
   \label{F}
   \Tr g[(\vec \sigma\cdot{\bf a}),(\vec \sigma\cdot{\bf b})]
   \equiv\\
   \frac 1 2\sum_{\epsilon_{1(2)}
   =\pm 1}\left(1+\epsilon_1\epsilon_2\frac{{\bf a}\cdot
   {\bf b}}{|{\bf a}||{\bf b}|}\right)
   g[\epsilon_1|{\bf a}|,\epsilon_2|{\bf b}|],
\end{multline}
where $g[x,y]$ denotes an analytical function (\eqref{F} then is proven via
series expansion) we can rewrite the noise power \eqref{noise} in
the form\cite{conditions of applicability}
\begin{gather}
   \label{S_ideal}
   S_{\alpha\beta}=
   S_{\alpha\beta}^{(a)}\sin^2\left(\frac{\theta_{\alpha\beta}}
   2\right)+
   S_{\alpha\beta}^{(p)}\cos^2\left(\frac{\theta_{\alpha\beta}}
   2\right),
\end{gather}
where $\alpha=3,5$, $\beta=4,6$ or vice versa. Here, $\theta_{\alpha\beta}$ denotes the angle between the magnetization
of leads $\alpha$ and $\beta$, e.g., $\cos(\theta_{56})={\bf a}\cdot{\bf b}$, and $\cos(\theta_{54})={\bf a}\cdot(-{\bf
b})$; below, we need configurations with different settings ${\bf a}$ and ${\bf b}$ and we define the angle
$\theta_{\bf ab}\equiv\theta_{56}$. The noise power for {\it antiparallel} (or {\it parallel}) orientations of the
ferromagnets $\alpha,\beta$ is denoted by $S_{\alpha\beta}^{(a(p))}$ [for example $S_{56}^{(p)}$ implies $\bf a
\upuparrows b$]. With these definitions, $F$ (see Eq.~\eqref{Bell_noise_F}) takes the from
\begin{gather}
   \label{F_S}
   F({\bf a},{\bf b})= -\cos(\theta_{{\bf ab}})
   \frac{S_{\alpha\beta}^{(a)}-S_{\alpha\beta}^{(p)}}
   {S_{\alpha\beta}^{(a)}+S_{\alpha\beta}^{(p)}}.
\end{gather}
The left hand side of Eq.\ \eqref{Bell_noise} has a maximum when
$\theta_{\bf ab}=\theta_{\bf a^\prime b}=\theta_{\bf a^\prime
b^\prime}=\pi/4$, and $\theta_{\bf ab^\prime}=3\theta_{\bf ab}$
(shown as in the photonic case\cite{book:Mandel_Wolf} with the
substitution $\theta \to \theta/2$). With this choice of angles the
Bell inequality \eqref{Bell_noise} with \eqref{F_S} reduces to
\begin{gather}
   \label{Bell-maximum}
   \left|\frac{S_{\alpha\beta}^{(a)}-S_{\alpha\beta}^{(p)}}
   {S_{\alpha\beta}^{(a)}+S_{\alpha\beta}^{(p)}}\right|\leq \frac
   {1}{\sqrt 2}.
\end{gather}

Consider then a biased superconductor (S) with grounded normal leads. The energy filters F$^{\rm e}_{1,2}$ (see
Fig.~1(b); we assume the filters to be perfectly efficient, i.e., $\Gamma_{1,2}\ll E_0$,
to begin with) select processes where Cooper pairs decay from S into different normal leads\cite{Lesovik1,Loss},
hence quasi-particle transmission between the leads is inhibited, $s^{\alpha\beta}_{ee}= s^{\alpha\beta}_{hh}=0$
for $\alpha$ even and $\beta$ odd. The trace in \eqref{noise} contains the Andreev processes $T_{he}^{\alpha\beta}
(1-T_{he}^{\alpha \beta})+ \{e\leftrightarrow h\}$, where $T_{ab}^{\alpha\beta}\equiv s_{ab}^{\alpha\beta}
(s^\dag)_{ba}^{\beta\alpha}$ (see also Ref. \onlinecite{Lesovik1}). Electrons and Andreev reflected holes
thus have opposite spin-polarization, hence $S^{(p)}=0$, and the Bell inequality \eqref{Bell-maximum} is
(maximally) violated, signalling that these beams are entangled.

Finally, we probe the robustness of our Bell test by allowing the filters $F^e_{1,2}$ to have  finite line widths
$\Gamma_{1,2}$. If, for instance $\Gamma_{1,2}\sim 2E_0$, the noise correlations will acquire a (small) $S^{(p)}$
contribution. According to \eqref{Bell-maximum} the BI can be violated even in this case, though not maximally;
alternatively, Eq. \eqref{Bell-maximum} can be used to estimate the quality of the filters F$^{\rm e}_{1,2}$. Here, we
have discussed the violation of BIs in an idealized situation ignoring paramagnetic impurities, spin-orbit interaction
{\it etc.} Imperfect filters should be considered in a similar way as in the quantum-optics
literature.\cite{book:Mandel_Wolf} Note that there are other inequalities which test entanglement for two-particle
\cite{Clauser-Horn} and for many-particle systems.\cite{Werner} The test of such inequalities can be implemented in a
similar manner as discussed above. Moreover, while electron-electron interactions were neglected here, it has been
suggested \cite{Loss_ee_interaction} that they do not destroy entanglement.

In conclusion, we propose a general form of BI-tests in solid-state systems formulated in terms of current-current
cross-correlators (noise), the natural observables in the stationary transport regime of a solid state device. For a
superconducting source injecting correlated pairs into a normal-metal fork completed with appropriate
filters,\cite{Lesovik1,Loss} the analysis of such BIs shows that this device constitutes a source of entangled
electrons when the fork is weakly coupled to the superconductor. Bell inequality-checks can thus be applied to test
electronic devices with applications in quantum communication and quantum computation where entangled states are basic
to their functionality.

We thank Yu.V.\ Nazarov and F.\ Marquardt for stimulating discussions. The research of N.M.C.\ and of G.B.L.\ was
supported by the RFBR, projects No.\ 00-02-16617, 01-02-06230, by Forschungszentrum J\"ulich (Landau Scholarship), by a
Netherlands NWO grant, by the Einstein center, and by the Swiss NSF.


\begin{thebibliography}{99}
\bibitem{lemma}
If $|X|=|Y|=1$ then $xy-xy^\prime=xy(1\pm x^\prime
y^\prime)-xy^\prime(1\pm x^\prime y)$. So $|xy-xy^\prime|\leq
|xy(1\pm x^\prime y^\prime)|+|xy^\prime(1\pm x^\prime y)|\leq(1\pm
x^\prime y^\prime)+(1\pm x^\prime y)=2\pm(x^\prime
y^\prime+x^\prime y)$. Thus $-(2+(x^\prime y^\prime+x^\prime
y))\leq xy-xy^\prime\leq 2-(x^\prime y^\prime+x^\prime y)$; the
last inequality is \eqref{lemma}.
\bibitem{note} If the particle source is a normal metal, then $\Lambda_\pm\sim \tau\omega_1^2g_1g_2$
can't be droped in \eqref{Bell_noise_F}, and \eqref{Bell_noise} can hardly be violated.
\bibitem{explanation}
This condition excludes processes where, for instance, an electron quasi-particle in lead 1 is not
absorbed by the terminals 3,5, but is reflected back to the superconductor and finally transformed
into a hole propagating through lead 2.
\end{thebibliography}
\end{document}